\documentclass{iopconfser}

\usepackage{graphicx}
\usepackage{wrapfig}
\usepackage{color}
\usepackage{epsfig}
\usepackage{float}
\usepackage{multirow}
\expandafter\let\csname equation*\endcsname\relax
\expandafter\let\csname endequation*\endcsname\relax
\usepackage{amsfonts}
\usepackage{amssymb}
\usepackage{amsmath}
\usepackage{tabularx,url,color}
\usepackage{hyperref}
\usepackage{ulem}
\usepackage{subfigure}
\usepackage{lineno}
\usepackage{upgreek}
\usepackage{comment}
\usepackage{xcolor}
\usepackage{tablefootnote}
\usepackage{lipsum}
\usepackage{tabularx}
\usepackage{lineno}
\usepackage{orcidlink}

\newcolumntype{M}[1]{>{\centering\arraybackslash}m{#1}}

\begin{document}

\title{From the Virgo interferometer calibration to the bias and uncertainty of the h(t) detector strain during the O4 run}

\author{Florian Aubin$^{1}$\orcidlink{0000-0003-1613-3142},Cervane Grimaud$^{2}$\orcidlink{0000-0001-7736-7730}, Benoît Mours$^{1}$, Thierry Pradier$^{1}$\orcidlink{0000-0001-5501-0060}, Loïc Rolland$^{2}$\orcidlink{0000-0003-0589-9687}, Monica Seglar-Arroyo$^{3}$\orcidlink{0000-0001-8654-409X
}, Hans Van Haevermaet$^{4}$, Pierre Van Hove$^{1}$\orcidlink{0000-0002-2431-3381
}, Didier Verkindt$^{2}$\orcidlink{0000-0003-4344-7227}}

\affil{$^1$Université de Strasbourg, CNRS, IPHC UMR 7178, F-67000 Strasbourg, France}
\affil{$^2$Laboratoire d’Annecy de Physique des Particules (LAPP), Univ. Grenoble Alpes, Université Savoie Mont Blanc, CNRS/IN2P3, F-74941 Annecy, France}
\affil{$^3$Institut de Fisica d’Altes Energies (IFAE), Barcelona Institute of Science and Technology, and ICREA, E-08193 Barcelona, Spain}
\affil{$^4$Universiteit Antwerpen, Prinsstraat 13, 2000 Antwerpen, Belgium}

\email{cervane.grimaud@lapp.in2p3.fr}
\begin{abstract}
  Since the first gravitational wave detection in 2015, ground-based interferometer sensitivities have significantly improved, requiring highly precise calibration to ensure accurate reconstruction of the h(t) strain signal. 
  In this talk we will outline the Virgo interferometer calibration steps performed in preparation of the O4b run started in April 2024. We will first describe the Photon Calibrator power devices intercalibration allowing for a 0.48\% precision on mirror displacement. Before explaining how the Photon Calibrator is used to calibrate every Virgo mirror actuators.
  We will also discuss the monitoring of the h(t) strain reconstruction during the run showing that, on the 10 Hz to 2~kHz band, the reconstructed strain achieves a precision of 2\% in modulus and 30 mrad in phase. Special emphasis will be given on the newly developed frequency-dependent bias and uncertainty computation method and the resulting online unbiasing of the h(t) strain. 
\end{abstract}

\section*{Introduction}
The existence of gravitational waves was predicted in 1916 by A. Einstein as a consequence of the general relativity theory \cite{Einstein:GR}. They are defined as perturbations of the spacetime metric propagating at the speed of light and emitted by accelerating compact objects.
The detection of gravitational waves is currently performed with a worldwide network of four interferometers managed by the LIGO-Virgo-KAGRA (LVK) collaborations \cite{Abbott_2020}.

This network is composed of ground interferometers using laser interferometry to measure tiny length variations between suspended test masses (mirrors) in the frequency band 10~Hz to 10~kHz \cite{Acernese_2015,TDR:ADV+}. The interferometer output is the dark fringe of the interference pattern and is used to reconstruct the detector strain, $h(t)$, that contains the gravitational wave information. The $h(t)$ strain amplitude is inversely proportional to the distance between the source and the observer. As such, a typical length variation caused by a gravitational wave going through one of the network interferometers is of the order of $10^{-18}$~m. Thus, to be able to detect gravitational waves with an interferometer, the test masses have to be isolated from external noises like seismic noise and their positions have to be very precisely controlled to keep the interferometer in its observing configuration. In Virgo, the different mirror positions are controlled with feedback loops using electromagnetic actuators (EM) (magnets placed at the back of the mirrors with coils in front of them). As such, the $h(t)$ detector strain needs to be reconstructed from the dark fringe signal taking into account all the control signals contributions and using the different mirrors actuators frequency response models. Those frequency responses are given by calibration measurements performed using another independent actuator as reference of mirror displacement.

\section{Virgo calibration}
For the Virgo interferometer, there are two actuators used as reference of mirror displacement. First, the Newtonian Calibrator (NCal) which uses a variable gravitation field produced by rotating masses to induce a known motion of a suspended mirror \cite{NCal:O3,NCal:O4}. Second, the Photon Calibrator (PCal) which uses the radiation pressure of an auxiliary laser beam sent towards the center of a mirror to induce a known displacement \cite{bib:2020_PCalO3,PhdLagabbe,PhdGrimaud}. During O4, the absolute reference of mirror displacement was the Newtonian Calibrator because of its displacement uncertainty of 0.12\%. However, this actuator can induce displacement only up to 150 Hz and only for one chosen frequency at a time. As such, the Photon Calibrator is used as a complement reference for the 10~Hz to 2~kHz observation band with an uncertainty of 0.48\%. The $\Delta x$ mirror displacement induced by the PCal laser on a test mass along the beam axis can be estimated using:
\begin{equation}
  \begin{aligned}
    \Delta x_{PCal}(f) &= A^{mech}_{PCal}(f) \; \frac{2 \cos(\theta)}{c} \; \frac{P_{Rx}}{(1-l_{vp})(1-l_M)}\; . \;S^{-1}_{PCal}(f) \equiv A_{PCal} \times P_{Rx} \\
  \end{aligned}
\end{equation}
Where $A^{mech}_{PCal}(f)$ is the mechanical response model of the mirror, $\theta$ is the angle of incidence of the PCal laser on the mirror, $P_{Rx}$ is the reflected laser beam power measured by a power sensor (integrating sphere) named Rx, $l_{vp}$ and $l_{M}$ are optical losses coefficients and $S_{PCal}(f)$ is the Rx sensor data acquisition chain sensing response \cite{PhdGrimaud,Intercalib_meas_note,Power_calib_note}.

Using the PCal as reference of mirror displacement, it is then possible to calibrate the electromagnetic actuators used for the interferometer control system. The calibration is based on the comparison of the detector's response $R$ to two different injection paths. First an injection $I_{ref}$ applied with a calibrated actuator of reference with a response model $A_{ref}$ and second an injection $I_{new}$ applied with the actuator to calibrate $A_{new}$. The output signal of the interferometer $O$ can be written as a combination of the injection signal, the actuator model and the interferometer response, $O=I_i \times A_i \times R$. Assuming R is the same during both data sets we can compute two transfer functions that can be combined to extract the actuator response of the system we want to calibrate. We get: 
\begin{equation}
  A_{new} = \left(\frac{O}{I_{new}}\right) .  \left(\frac{O}{I_{ref}}\right)^{-1} \times A_{ref}
\end{equation}
This is the general method for actuator calibration that we apply to all mirror actuators \cite{CALI_O3,PhdGrimaud}.

\section{h(t) detector strain reconstruction principle}
  
The $h(t)$ strain reconstruction is performed by subtracting, from the dark fringe signal, the contribution of each longitudinal control signal using the different mirror actuator responses computed thanks to the calibration measurements presented above. For Virgo, the algorithm used to reconstruct the detector strain is called \textit{Hrec}.

This process needs to be monitored to assess if the reconstruction is working properly and to provide the reconstruction bias and uncertainty. This is done using a measurement called $h_{raw} / h_{inj}$, where the transfer function between the reconstructed strain ($h_{raw}$) and the strain computed using only the calibration model of the used actuator ($h_{inj}$) is computed. If Hrec is working perfectly the result of this transfer function should be 1 for the modulus and 0 for the phase, deviations from those values give an information on the reconstruction bias and uncertainty. There are two different types of measurements used to monitor the Hrec bias. The first type corresponds to the Permanent Lines: a set of 16 lines (sinusoidal signals) permanently injected with frequencies going from 16~Hz to 1~kHz. The second type refers to the Weekly Lines: a set of 27 lines injected for one minute once a week during a dedicated calibration slot, with frequencies from 18~Hz to 1158~Hz. The permanent lines and weekly lines are injected using both the EM and PCal actuators.

\section{h(t) reconstruction bias and uncertainty computation method}
The goal for O4 was to unbias directly the reconstructed strain inside the Hrec algorithm, and to compute a frequency-dependent reconstruction uncertainty. To do this, a method was developed to compute the reconstruction bias using the weekly lines, and to compute the reconstruction uncertainty using the permanent lines.
The computation method is the same for both the bias and uncertainty computations and uses either the weekly or permanent lines data points and takes an average of the modulus and phase values over a given time period (typically one month). Those mean values and error values are used to compute a Gaussian distribution of the data. Then, a random selection of $N$ points per lines (typically 1000) is performed inside the modulus and phase distributions of all the 27 frequency lines.
With the $N$ randomly selected sets of data, a linear interpolation between the 27 points is done, computing a value of modulus and phase every 0.125 Hz frequency bin from the 10~Hz to 10~kHz frequency range.
For the low frequencies, before the first weekly line frequency, the first interpolation coefficient computed with the closest two lines is used to compute the extrapolated value up to the 10~Hz bin. At high frequency, the interferometer is not controlled, so the last frequency bin (at 10 kHz) is set to 1 for the modulus and 0 for the phase.

Finally, we get $N$ sets of interpolated data points, which means that we get an $N$ points distribution for each frequency bin. From those distributions we can compute a mean and a standard deviation, which represent the frequency-dependent bias and uncertainty on the $h(t)$ modulus and phase.

\section{Bias and uncertainty results during the O4 run}

\begin{figure}[h!t]
  \begin{center}
    \subfigure[\label{subfig:ChekHrec_online}$h_{raw}/h_{inj}$]{
      \includegraphics[angle=0,width=0.47\linewidth]{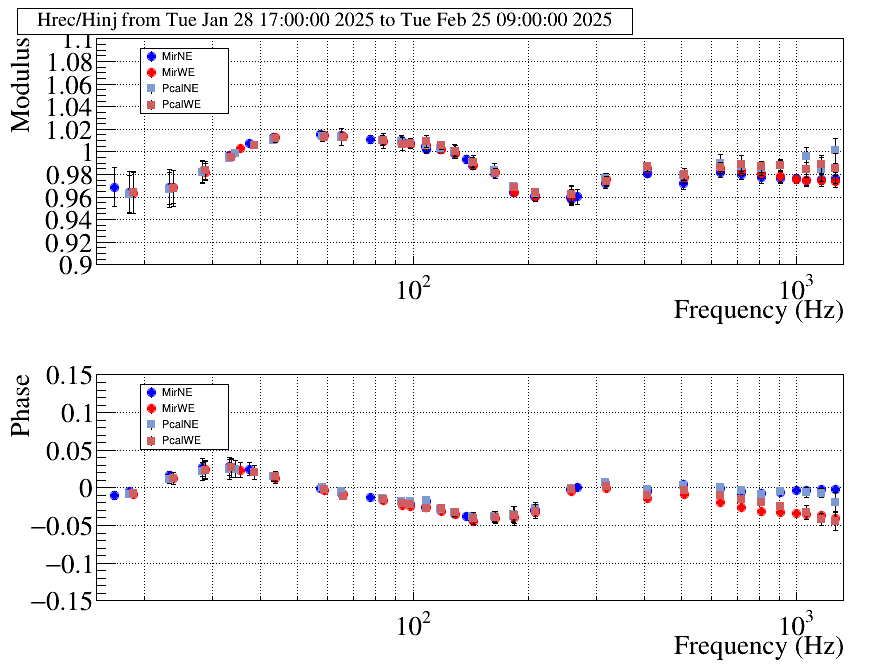}}
    \subfigure[\label{subfig:ChekHrec_unbias}$h_{unbias}/h_{inj}$]{
      \includegraphics[angle=0,width=0.47\linewidth]{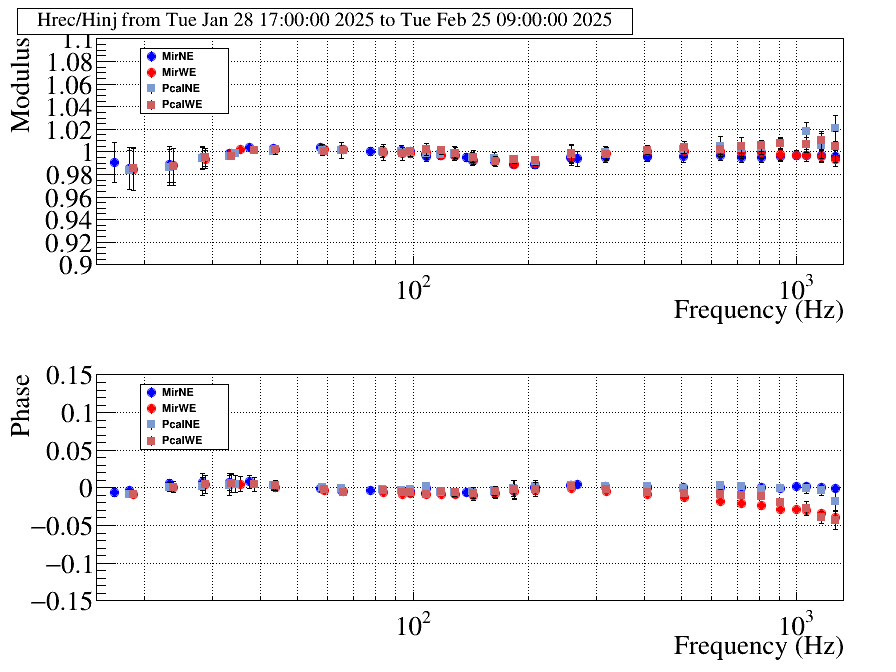}}
      \caption{ \label{fig:CheckHrec_comparison} Comparison between the $h_{raw}/h_{inj}$ measurements performed using an average of the weekly lines data from the 28th of January to the 25th of February 2025 and the $h_{unbias}/h_{inj}$ measurements performed with the same average period.}
  \end{center}
\end{figure}

Following this computation method and using the weekly injections we are able to compute a bias estimation which is then used to correct online the reconstructed strain $h_{raw}$ to get the unbiased one $h_{unbias}$. We can monitor the unbiasing by doing an $h_{unbias}/h_{inj}$ measurement which will show the residual bias of the reconstruction. This allows to see if the unbiasing is working well or if we need to update the reconstruction bias. In figure~\ref{fig:CheckHrec_comparison} we see that the unbiasing allows to get a reconstructed strain with a residual bias within less than 2\% over 10 Hz to 1 kHz and less than 1\% over 30 to 500 Hz for the modulus. For the phase the residual bias is less than 15 mrad from 10 to 1 kHz. 

\begin{wrapfigure}{R}{0.49\textwidth}

  \begin{center}
      \includegraphics[width=0.48\textwidth]{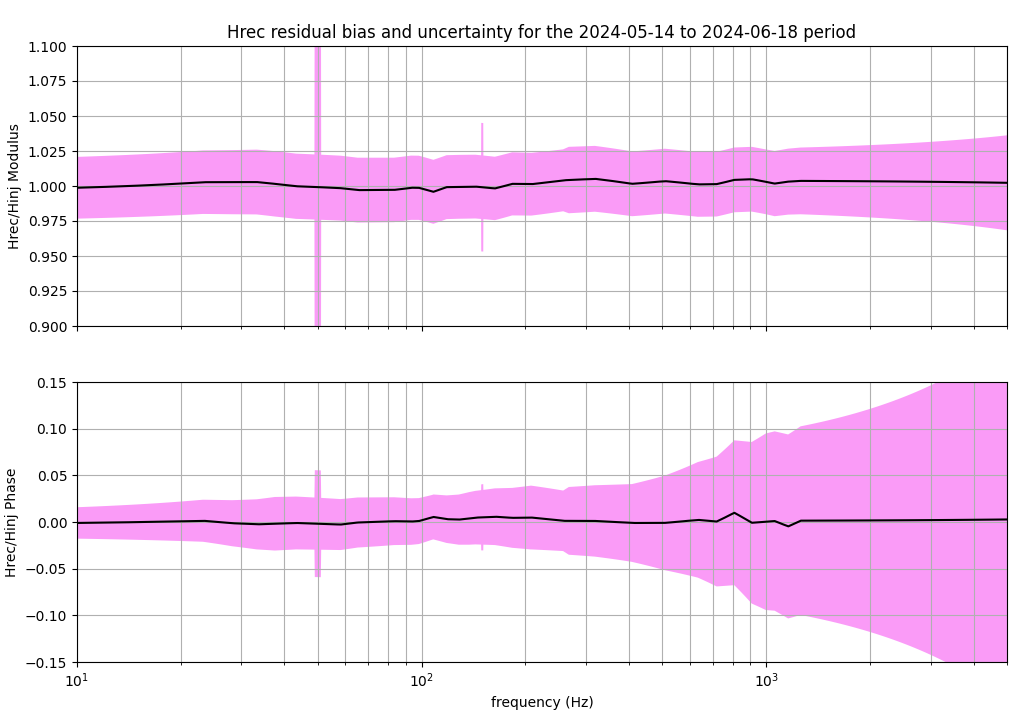}%
  \end{center}
  
  \caption{Result of the reconstruction residual bias and uncertainty for the period 2024-05-14 to 2024-06-18. The residual bias is computed using the weekly lines and the uncertainty is computed using the permanent lines.}
  \label{fig:HrecUncertainty}
\end{wrapfigure}%

The goal for O4 was to be able to compute a frequency-dependent reconstruction uncertainty. This is performed using the exact same method as for the bias computation but using the unbiased strain ($h_{unbias}$) instead of the raw strain ($h_{raw}$) and using the permanent lines instead of the weekly lines. The idea behind using the permanent lines is that since they are permanently injected, they allow accounting for the systematic uncertainties coming from variations in $h(t)$ reconstruction, due to interferometer alignment variations or optical response variations, that have a short timescale (lower than 1 hour). The uncertainties from the calibration are added quadratically to the uncertainties estimated with this method. This includes a conservative estimation of the PCal uncertainty of 0.6\% made before the start of the run and 0.5\% from the actuators models for the modulus and 5~mrad on the phase coming from the actuators models. There is also additional errors around 50 Hz and 150 Hz, which corresponds to the power supply main frequency and its second harmonics, that were added after looking at broadband injections used to compute $h_{rec}/h_{inj}$ \cite{PhdGrimaud}. 

The final step, is to combine the residual bias computed with the weekly lines and the uncertainty at $\pm 1\sigma$ computed with the permanent lines, for both the modulus and the phase of $h_{unbias}/h_{inj}$. This process is done for monthly data chunks. The result obtained for the chunk from the 14th of May to the 18th of June 2024 is presented in figure~\ref{fig:HrecUncertainty}.

\section*{Conclusion}
In conclusion, the $h(t)$ strain is reconstructed and unbiased online with a latency of ${\sim}10$~s before being distributed to the LVK low-latency data analysis pipelines. The {\it online} $h(t)$ strain contains a preliminary frequency-dependent uncertainty estimated using the permanent lines injections from before the start of the run. 
In addition, an {\it offline} $h(t)$ strain is released on a monthly basis with updated residual bias and uncertainties computed using the permanent lines injection of the month (see figure~\ref{fig:HrecUncertainty}) and with $h(t)$ set to zero for periods of bad quality. Using a new bias and uncertainty computation method, we were able to unbias online the reconstructed strain. This was not done during O3 and has been setup for Virgo during O4. In addition, we were able to improve the uncertainty computation method which allows to compute a conservative frequency-dependent reconstruction uncertainty over the 10~Hz to 2~kHz frequency band instead of the frequency-independent uncertainty used during O3. The residual reconstruction bias and the uncertainties are computed on a monthly basis allowing to monitor their evolution. We were able to see that the bias is very stable in time, and it was changed only twice over 16 months. The overall strain reconstruction uncertainty is below 2.5\% in modulus in the 10~Hz to 1~kHz band and within 25~mrad in the phase below 200 Hz before increasing up to 100 mrad at 1~kHz.

\section*{Acknowledgments}
The authors gratefully acknowledge the Italian Istituto Nazionale di Fisica Nucleare (INFN),  
the French Centre National de la Recherche Scientifique (CNRS) and
the Netherlands Organization for Scientific Research (NWO), 
for the construction and operation of the Virgo detector
and the creation and support of the EGO consortium.
The authors also gratefully acknowledge research support from these agencies as well as by 
the Spanish  Agencia Estatal de Investigaci\'on, 
the Consellera d'Innovaci\'o, Universitats, Ci\`encia i Societat Digital de la Generalitat Valenciana and
the CERCA Programme Generalitat de Catalunya, Spain,
the National Science Centre of Poland and the European Union – European Regional Development Fund; Foundation for Polish Science (FNP),
the Hungarian Scientific Research Fund (OTKA),
the French Lyon Institute of Origins (LIO),
the Belgian Fonds de la Recherche Scientifique (FRS-FNRS), 
Actions de Recherche Concertées (ARC) and
Fonds Wetenschappelijk Onderzoek – Vlaanderen (FWO), Belgium,
the European Commission.
The authors gratefully acknowledges support from the French Agence Nationale de la Recherche for the project ACALCO (ANR-21-CE31- 0024).
The authors gratefully acknowledge the support of the NSF, STFC, INFN, CNRS and Nikhef for provision of computational resources.

\bibliographystyle{unsrt}
\bibliography{biblio.bib}

\end{document}